\documentclass[aps,prb,twocolumn,showpacs,floatfix]{revtex4}
\usepackage{verbatim} 
\usepackage{amsmath,amssymb,graphicx}
\usepackage{epsfig}
\usepackage{graphicx}
\usepackage{enumerate}
\usepackage[toc,page,header]{appendix}

\newcommand{\beq}{\begin{equation}}
\newcommand{\eeq}{\end{equation}}

\newcommand{\bea}{\begin{eqnarray}}
\newcommand{\eea}{\end{eqnarray}}

\begin{document}


\title{
The probe technique far-from-equilibrium:\\
Magnetic field symmetries of nonlinear transport}
\author{Salil Bedkihal$^1$, Malay Bandyopadhyay$^2$, Dvira Segal$^1$ }
\affiliation{$^1$Chemical Physics Theory Group, Department of
Chemistry, University of Toronto, 80 Saint George St. Toronto,
Ontario, Canada M5S 3H6}
\address{$^2$
School of Basic Sciences, Indian Institute of Technology Bhubaneswar, 751007, India}


\begin{abstract}
The probe technique is a simple mean to incorporate elastic and inelastic processes
into quantum dynamics.
Using numerical simulations, we demonstrate that this tool can be employed
beyond the analytically tractable linear response regime, providing a stable solution
for the  probe parameters: temperature and chemical potential.
Adopting four probes: dephasing, voltage, temperature, and voltage-temperature, mimicking
different elastic and inelastic effects, we focus on 
magnetic field and gate voltage symmetries of charge current and heat current in
Aharonov-Bohm interferometers, 
potentially far-from-equilibrium. Considering
electron current, we prove analytically that in the linear
response regime inelastic scattering processes do not break the Onsager symmetry. 
Beyond linear response,
even (odd) conductance terms obey an odd (even) symmetry
with the threading magnetic flux, as long as the system acquires a spatial inversion symmetry.
When spatial asymmetry is introduced 
particle-hole symmetry assures that nonlinear conductance terms maintain certain symmetries with
respect to magnetic field and gate voltage.
These analytic results are supported by numerical simulations.
Analogous results are obtained for the electron heat current.
We also demonstrate that
a double-dot Aharonov-Bohm interferometer acts as a rectifier
when two conditions are 
met: (i) many-body effects are included, here in the form of
inelastic scattering, and (ii) time reversal symmetry is broken.
\end{abstract}

\date{\today}

\maketitle

\section{Introduction}
\label{intro}

\begin{figure}[t]
\hspace{-16mm} {\hbox{\epsfxsize=200mm \epsffile{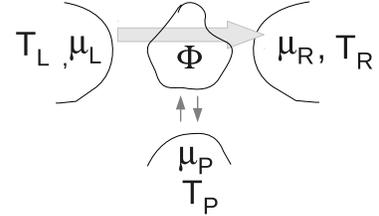}}}
\vspace{-107mm} \hspace{-2mm} \caption{Scheme of our setup. The
horizontal arrow stands for the charge current across the conductor.
The two parallel arrows represent currents
into and from the $P$ terminal, serving to induce elastic, 
inelastic, and dissipative effects. The parameters of the probe $\mu_P$
and $T_P$ are determined in a self consistent manner, to satisfy the
respective probe condition.} \label{FigS}
\end{figure}

Phase-breaking and energy dissipation processes arise due to the
interaction of electrons with other degrees of freedom, e.g., with
electrons, phonons, and defects. While  
an understanding of such
effects, from first principles, is the desired objective of 
numerous computational approaches \cite{Review}, 
simple analytical treatments are advantageous as they
allow one to gain insights into transport phenomenology.
The markovian quantum master equation and its variants (Lindblad,
Redfield) is simple to study and interpret \cite{VonKampen}, and as
such it has been extensively adopted in studies of charge, spin, exciton, and
heat transport. 
It can be derived   systematically, from projection operator techniques \cite{Qopen},
and phenomenologically by
introducing damping terms into 
the matrix elements of the reduced density matrix, to include dephasing and
inelastic processes into the otherwise  coherent dynamics.

B\"uttiker's probe technique \cite{Buttiker,Pastawski,Ando} and
its modern extensions to thermoelectric problems
\cite{Jacquet,Casati1,Casati2,Seifert}, atomic-level thermometry \cite{Stafford},
and beyond linear response situations \cite{Malay,Ming,Unique,FPU} procure an alternative
route for introducing decoherence and inelastic processes into
coherent conductors. 
The probe is an electronic component  \cite{Vexp}, and it allows one to obtain information about
local variables, chemical potential and temperature, deep within the
conductor. 
When coupled strongly to the system, the probe can
alter intrinsic transport mechanisms.

Probes can be constructed to induce distinct effects: {\it Elastic dephasing}
processes are implemented by incorporating a ``dephasing`` probe,
enforcing the requirement that the net charge current towards the
probe terminal, at any given energy, vanishes
\cite{Dprobe1,Dprobe2}. Inelastic {\it heat dissipative} effects
are included by a voltage probe, by demanding that
the total-net charge current to the probe terminal nullifies
\cite{Buttiker,Pastawski,Ando}. This process dissipates heat
since electrons leaving the system to the probe re-enter the
conductor after being thermalized. In the complementary temperature
probe {\it charge dissipation} is allowed at the probe, but the
probe temperature is tuned such that the net heat current at the probe is annulled
 \cite{Anderson}. The voltage-temperature probe,
also referred to as a ``thermometer", requires both charge current and heat
current at the probe to vanish. In this case inelastic  - energy exchange effects are allowed on the
probe, but heat dissipation and charge dissipation effects are excluded.

The probe technique has been used in different
applications, particularly for the exploration of the ballistic to
diffusive (Ohm's law and Fourier's law) crossover in electronic \cite{Pastawski,Dhar1} and
phononic conductors \cite{Lebo,Roy1,Roy2}.
More recently, the effect of thermal rectification has been studied
in phononic systems by utilizing the temperature probe as a mean to
incorporate effective anharmonicity \cite{LangC,Ming,Malay}. A
full-counting statistics analysis of conductors including dephasing
and voltage probes has been carried out in Ref. \cite{ButtikerFCS}
%
The probe parameters,  temperature and
chemical potential, can be derived analytically 
when the conductor is set close to equilibrium
\cite{Buttiker,Vexp,Jacquet}. 
Far-from-equilibrium, while these parameters can be technically defined and their 
uniqueness \cite{Unique} allows for a physical interpretation, 
an exact analytic solution is missing. 
However, 
recent studies have demonstrated that iterative numerical schemes can reach
a stable solution for the temperature probe \cite{Malay,FPU}. These techniques have 
been then used for following phononic heat transfer in the deep quantum limit, 
far-from-equilibrium  \cite{Malay,FPU}.

The Onsager-Casimir symmetry relations \cite{OnsagerC} are satisfied
in phase-coherent conductors, reflecting the microreversibility of
the scattering matrix. In Aharonov-Bohm interferometers with
conserved electron current this symmetry is displayed  by the
``phase rigidity" of the (linear) conductance oscillations with the
magnetic field $B$, $G_1(B)=G_1(-B)$ \cite{Imry,Yacoby}. Beyond
linear response, the phase symmetry of the conductance is not
enforced, and several experiments
\cite{Linke1,Linke2,breakE1,breakE2,breakE3,breakE4,breakE5,Hernandez11} have
demonstrated its breakdown. Supporting theoretical works have
elucidated the role of many-body interactions in the system 
\cite{break1, break2,
break3, Kubo, Meir}, typically approaching the problem by
calculating the screening potential within the conductor in a self-consistent manner,
a procedure often limited to low-order conduction
terms \cite{break1, break3, Kubo}.

The present manuscript is focused on the application of the probe technique
to quantum open system problems, possibly in far-from-equilibrium situations.
We adopt different probes and
consider the role of
elastic dephasing, heat dissipation, and charge dissipation
processes on magnetic field, temperature bias, and voltage bias
symmetries of charge current, rectification, and heat current in
Aharonov-Bohm (AB) interferometers.
Our main objective is the development, and analysis of the breakdown,
of symmetry relations for nonlinear transport beyond the Onsager-Casimir
limit, in the presence incoherent effects.

This work extends our recent study \cite{Salil3} in several ways:
(i) We consider four different
probes (dephasing, voltage, temperature and voltage-temperature) and
demonstrate with numerical simulations a stable 
solution for the probe parameters and a facile convergence, far-from-equilibrium.
(ii) We then consider a generic model for an AB interferometer, 
susceptible to elastic and inelastic effects, and
study its conductance behavior: 
(ii.a) We provide a detailed proof for
the validity of phase rigidity under the voltage probe
in the linear response regime. (ii.b) We prove the development of
new-general set of magnetic-field symmetry relations away from linear response when the
device is geometrically symmetric.  (ii.c) 
These magnetic-field symmetries are invalidated under spatial asymmetry, but we prove that
generalized magnetic field-gate voltage symmetry relations are obeyed, a result of particle-hole symmetry.
(ii.d) We discuss the operation of the 
 double-dot interferometer, susceptible to inelastic effects, 
as a charge rectifier, when time reversal symmetry is broken.

The paper is organized as follows. In Sec. \ref{formalism} we
provide expressions for the charge and heat currents in the Landauer
formalism and discuss four types of probes, inducing different
effects. Sec. \ref{measure} introduces the main observables of
interest and summarizes our principal results. 
Sec. \ref{dephasing} covers setups that fulfill phase rigidity. 
Magnetic field symmetry
relations in a spatially symmetric setup are derived in
Sec. \ref{symmetry}.
Magnetic field- gate voltage symmetries, valid for generic
double-dot AB interferometer models, are  presented in Sec. \ref{violation}.
Supporting numerical simulations are included in Sec. \ref{simulation}.
Sec. \ref{summary} concludes. For simplicity, we set $e$=1, $\hbar=1$ and
$k_B=1$. The words ``diode" and ``rectifier" are used
interchangeably in this work, referring to a dc-rectifier.

\section{Formalism}
\label{formalism}

In the scattering formalism of Landauer and B\"uttiker
\cite{Landauer,Buttiker,Buttiker-4,Datta} interactions between
particles are neglected. One can then express the charge current
from the $\nu$ to the $\xi$ terminal in terms of the transmission
probability $\mathcal T_{\nu,\xi}(\epsilon)$, a function which
depends on the energy of the incident electron,
\bea
I_{\nu}(\phi)&=&\int_{-\infty}^{\infty}d\epsilon \Big[\sum_{\xi \neq \nu}
\mathcal T_{\nu,\xi}(\epsilon,\phi)
f_{\nu}(\epsilon)
-\sum_{\xi \neq \nu}\mathcal T_{\xi,\nu}(\epsilon,\phi)
f_{\xi}(\epsilon) \Big]. \nonumber\\ \label{eq:curr} \eea
The magnetic field is
introduced via an Aharonov-Bohm flux $\Phi$ applied through the conductor,
with the magnetic phase $\phi=2\pi\Phi/\Phi_0$, $\Phi_0=h/e$ is the magnetic flux quantum.
The transmission function can be written in terms of the Green's
function of the system and the self energy matrices. Explicit
expressions for a particular model are included in Sec. \ref{violation}. The
Fermi-Dirac distribution function
$f_{\nu}(\epsilon)=[e^{\beta_{\nu}(\epsilon-\mu_{\nu})}+1]^{-1}$ is
defined in terms of the chemical potential $\mu_{\nu}$ and the
inverse temperature $\beta_{\nu}$.

Our analysis below relies on two basic relations. First, the
transmission coefficient from the $\xi$ to the $\nu$ reservoir
obeys reciprocity, given the unitarity and time reversal symmetry of the scattering matrix,
\bea
\mathcal T_{\xi,\nu}(\epsilon,\phi)= \mathcal T_{\nu,\xi}(\epsilon,-\phi).
\label{eq:R1}
\eea
Second, the total probability is conserved,
\bea
\sum_{\xi\neq \nu} \mathcal T_{\xi,\nu}(\epsilon,\phi)=
\sum_{\xi\neq \nu} \mathcal T_{\nu,\xi}(\epsilon,\phi).
\label{eq:R2}
\eea
A proof for the second relation, in the presence of a probe, is included in Appendix A of Ref.
\cite{Kubo}, based on the Green's function formalism.

In this work we consider a setup including three terminals,
$L$, $R$ and $P$, where the $P$ terminal serves as the probe,
see Fig. \ref{FigS}. We focus below on the steady-state
charge current from the $L$ reservoir to the central system ($I_L$) and from the
probe to the system ($I_P$),
\bea
I_L(\phi)&=&
\int_{-\infty}^{\infty} d\epsilon \Big[ \mathcal T_{L,R}(\epsilon,\phi)f_L(\epsilon)-
 \mathcal T_{R,L}(\epsilon,\phi)f_R(\epsilon)
\nonumber\\
&+& \mathcal T_{L,P}(\epsilon,\phi)f_L(\epsilon)-
 \mathcal T_{P,L}(\epsilon,\phi)f_P(\epsilon,\phi)
\Big],
\label{eq:IL}
\eea
%
%
\bea I_P(\phi)&=& \int_{-\infty}^{\infty}d\epsilon \Big[\mathcal
T_{P,L}(\epsilon,\phi)f_P(\epsilon,\phi)
- \mathcal
T_{L,P}(\epsilon,\phi)f_L(\epsilon)
\nonumber\\
&+&
\mathcal T_{P,R}(\epsilon,\phi)f_P(\epsilon,\phi) - \mathcal
T_{R,P}(\epsilon,\phi)f_R(\epsilon) \Big].
\label{eq:IP} \eea
Similarly, we can write the heat current at the $\nu=L$ terminal as
\bea &&Q_L(\phi)= \int_{-\infty}^{\infty} d\epsilon
\left(\epsilon-\mu_{L}\right)\Big[ \mathcal
T_{L,R}(\epsilon,\phi)f_L(\epsilon)
\nonumber\\
&&-
 \mathcal T_{R,L}(\epsilon,\phi)f_R(\epsilon)
+ \mathcal T_{L,P}(\epsilon,\phi)f_L(\epsilon)-
 \mathcal T_{P,L}(\epsilon,\phi)f_P(\epsilon,\phi)
\Big]. \nonumber\\ \label{eq:QL} \eea
The probe heat current follows an analogous form.
%
%
The probe distribution function is determined by the probe condition.
It is generally affected by the magnetic
flux, as we demonstrate in Secs. \ref{dephasing}-\ref{simulation}. 

For convenience, we simplify next our notation. First, we drop the
reference to the energy of incoming electrons $\epsilon$ in both
transmission functions and distribution functions. Second, since all
integrals are evaluated between $\pm \infty$, we do not put the
limits explicitly. Third, unless otherwise mentioned
 $f_ P$, $\mu_P$ and all transmission coefficients are
evaluated at the phase $+\phi$, thus we do not explicitly write the
phase variable. If we do need to consider e.g. the transmission function
$\mathcal T_{\nu,\xi}(-\phi)$, we write instead the complementary
expression, $\mathcal T_{\xi,\nu}(\phi)$.

{\it Dephasing probe.} We implement elastic dephasing effects
by demanding that 
the energy-resolved particle current diminishes in the probe,
\bea I_P(\epsilon)=0 \,\,\,\,\, {\rm with} \,\,\,\ I_P=\int
I_P(\epsilon)d\epsilon. \label{eq:Dprobe} \eea
Using this condition, Eq. (\ref{eq:IP}) provides a closed form for the corresponding
(flux-dependent) probe distribution, not necessarily in the
form of a Fermi function.

{\it Voltage probe.} We introduce dissipative inelastic
effects into the conductor using the voltage probe technique. 
The three reservoirs are maintained at the same inverse temperature $\beta_a$, but
the $L$ and $R$ chemical potentials are made distinct, $\mu_L\neq \mu_R$. Our
objective is to obtain $\mu_P$, and it is reached by
demanding that the net-total particle current flowing into the
$P$ reservoir diminishes,
\bea
I_P=0.
\label{eq:Vprobe}
\eea
This choice allows for dissipative energy exchange processes to
take place within the probe. In the linear response regime Eq. (\ref{eq:IP})
can be used to derive an
analytic expression for $\mu_P$.
In far-from-equilibrium situations we obtain the unique \cite{Unique}
chemical potential of the probe numerically, 
using the Newton-Raphson method \cite{NR}
\bea \mu_P^{(k+1)}= \mu_P^{(k)} - I_P(\mu_P^{(k)})
\left[\frac{\partial I_P(\mu_P^{(k)})}{\partial \mu_P}\right]^{-1}.
\label{eq:NR} \eea
The current $I_P(\mu_P^{(k)})$ and its derivative are evaluated from Eq. (\ref{eq:IP}) using the probe
(Fermi) distribution with $\mu_P^{(k)}$. 
Note that the self-consistent probe solution varies with the magnetic
phase.

{\it Temperature probe.} In this scenario the three reservoirs $L$,$R$,$P$ are
maintained at the same chemical potential $\mu_a$, but the temperature at
the $L$ and $R$ terminals vary,
$T_L\neq T_R$. 
The probe temperature $T_P=\beta_P^{-1}$ is determined 
by requiring the net heat current at the probe to satisfy
\bea Q_P=0. \label{eq:Tprobe} \eea
This constraint allows for charge dissipation into the probe since
we do not require Eq. (\ref{eq:Vprobe}) to hold. We can obtain the
temperature $T_P$ numerically by following an iterative procedure,
\bea
T_P^{(k+1)}= T_P^{(k)} - Q_P(T_P^{(k)})  \left[\frac{\partial Q_P(T_P^{(k)})}{\partial T_P}\right]^{-1}.
\label{eq:TNR}
\eea
The probe temperature depends on the flux $\phi$, see Appendix B.

{\it Voltage-temperature probe.} This probe acts
as an electron thermometer at weak coupling.
We set the temperatures $T_{L,R}$ and the potentials $\mu_{L,R}$,
and demand that 
\bea
I_P=0 \,\,\,\ {\rm and} \,\, Q_P=0.
\label{eq:TVprobe}
\eea
In other words, the charge and heat currents in the conductor satisfy
$I_L=-I_R$ and $Q_L=-Q_R$, since neither charge nor heat are allowed to dissipate
at the probe. Analytic results can be obtained in the linear
response regime, see for example Refs. \cite{Dhar1,Jacquet}.
Beyond that  equation  (\ref{eq:TVprobe}) can be solved
self-consistently, to provide $T_P$ and $\mu_P$. This can be done by
utilizing the two-dimensional Newton-Raphson
method,
\bea
\mu_P^{(k+1)}&=& \mu_P^{(k)} - D^{-1}_{1,1} I_P(\mu_P^{(k)},T_P^{(k)})  - D^{-1}_{1,2} Q_P(\mu_P^{(k)},T_P^{(k)})
\nonumber\\
T_P^{(k+1)}&=& T_P^{(k)} -  D^{-1}_{2,1} I_P(\mu_P^{(k)},T_P^{(k)})  - D^{-1}_{2,2} Q_P(\mu_P^{(k)},T_P^{(k)}),
\nonumber\\
\label{eq:TVprobes}
\eea
where the Jacobean  $D$ is re-evaluated at every iteration,
\[D(\mu_P,T_P)\equiv \left( \begin{array}{cc}
\frac{\partial I_P(\mu_P,T_P)}{\partial \mu_P} & \frac{\partial I_P(\mu_P,T_P)}{\partial T_P}   \\
\frac{\partial Q_P(\mu_P,T_P)}{\partial \mu_P}  & \frac{\partial Q_P(\mu_P,T_P)}{\partial T_P}
\end{array} \right)\]
%
We emphasize that besides the dephasing probe, the
function $f_P(\phi)$ is assumed to take the form of a Fermi-Dirac
distribution function.

\section{Symmetry measures and main results}
\label{measure}

In the main body of this paper we restrict ourselves to
voltage-biased junctions, $\mu_L\neq \mu_R$, while setting
$T_a=\beta_a^{-1}=T_L=T_R$. We also limit our focus to charge conserving systems
satisfying
\bea I(\phi)\equiv I_L(\phi)=-I_R(\phi), \label{eq:cc} \eea
and study the role of elastic dephasing (dephasing probe) and
dissipative (voltage probe) and non-dissipative (voltage-temperature
probe) inelastic effects on the charge transport symmetries with
magnetic flux. In Appendix B we complement this analysis by
considering a temperature-biased heat-conserving junction, $T_L\neq
T_R$, $\mu_a=\mu_L=\mu_R$ and $Q_L=-Q_R$. We then study the phase
symmetry of the heat current, allowing for  charge dissipation.
Thermoelectric effects are not considered in this work.

We now define several measures for quantifying phase symmetry
in a voltage-biased three-terminal junction satisfying
Eq. (\ref{eq:cc}). Expanding the charge current in powers of the
bias $\Delta \mu$ we write \cite{Persistent}
\bea
I(\phi)=G_1(\phi) \Delta\mu + G_2(\phi) (\Delta\mu)^2  + G_3(\phi)(\Delta\mu)^3 +...
\label{eq:II}
\eea
with $G_{n>1}$ as the nonlinear conductance coefficients.
In this work we study relations between two quantities: a measure for the
magnetic field asymmetry
\bea
\Delta I(\phi)\equiv \frac{1}{2}[I(\phi)-I(-\phi)],
\label{eq:DeltaIphi}
\eea
and the dc-rectification current,
\bea
\mathcal R(\phi)&\equiv& \frac{1}{2}[I(\phi)+\bar I(\phi)]
\nonumber\\
&
=& G_2(\phi) (\Delta\mu)^2  + G_4(\phi) (\Delta\mu)^4+ ...
\label{eq:R0}
\eea
with $\bar I$ defined as the current obtained upon interchanging
the chemical potentials of the two terminals.
We also study the behavior of odd conductance terms,
\bea
\mathcal D(\phi)&\equiv&
\frac{1}{2}[I(\phi)-\bar I(\phi)]
\nonumber\\
&=&
G_1(\phi) \Delta\mu  + G_3(\phi) (\Delta\mu)^3+ ...
\label{eq:D0}
\eea
For a non-interacting system we expect the relation
\bea
I(\phi)=-\bar I(-\phi)
\label{eq:noninte}
\eea
to hold. Combined with Eq. (\ref{eq:II}) we immediately note that
$G_{2n+1}(\phi)=G_{2n+1}(-\phi)$ and $G_{2n}(\phi)=-G_{2n}(-\phi)$
with $n$ as an integer.
We show below that these relations are obeyed in a
symmetric junction even when many-body interactions (inelastic
scattering) are included. This result is not trivial since the included many-body interactions are reflected
by probe parameters which depend on the applied bias in a nonlinear manner and the magnetic phase
in an asymmetric form, thus, we cannot assume Eq. (\ref{eq:noninte}) to immediately hold.

Using Eq. (\ref{eq:IL}), we express the deviation from the magnetic field symmetry as
\bea
\Delta I&=&\frac{1}{2}
\int \left[ \mathcal T_{L,R} - \mathcal T_{R,L}\right] (f_L+f_R)d\epsilon
\nonumber\\
&+& \frac{1}{2} \int \left[ \mathcal T_{L,P} - \mathcal T_{P,L}\right] f_L d\epsilon
\nonumber\\
&+& \frac{1}{2}
\int \left[ \mathcal T_{L,P}f_P(-\phi) - \mathcal T_{P,L} f_P(\phi)
\right] d\epsilon
\eea
We  use the probability conservation, Eq.  (\ref{eq:R2}),
and simplify this relation,
\bea
\Delta I&=& \frac{1}{2}
\int \left[ \mathcal T_{L,R} - \mathcal T_{R,L}\right] f_Rd\epsilon
\nonumber\\
&+& \frac{1}{2}
\int \left[ \mathcal T_{L,P}f_P(-\phi) - \mathcal T_{P,L} f_P(\phi)
\right]d\epsilon.
\label{eq:DeltaI}
\eea
The rectification current can be written as
\bea
\mathcal R= \int \frac{\mathcal T_{P,L}-\mathcal
T_{P,R}}{4}(f_L+f_R-f_P(\phi)-\bar f_P(\phi)) d\epsilon
\label{eq:R} \eea
with $\bar f_P$ as the probe distribution when the biases $\mu_L$ and
$\mu_R$ are interchanged.
%

Our results are organized by systematically departing from quantum coherent scenarios, 
the linear response regime, and spatially symmetric situations.
The paper includes four parts, and we now summarize our main
results:

(i) {\it Phase Rigidity.} In Sec. \ref{dephasing}  we discuss two scenarios that do obey
the Onsager-Casimir
symmetry relation $I(\phi)$=$I(-\phi)$: It is maintained in the presence
of elastic dephasing effects even beyond linear response. This relation is
also valid when inelastic scatterings are included, albeit only in the
linear response regime. While these results are not new
\cite{Buttiker-4}, we include this analysis here so as to clarify
the role of inelastic effects in breaking the Onsager symmetry,
beyond linear response. 

(ii) {\it Magnetic field (MF) symmetry relations beyond linear response.} 
In Sec. \ref{symmetry} we derive magnetic-field symmetry relations
that hold
far-from-equilibrium in {\it spatially symmetric} junctions susceptible to
inelastic effects, $\mathcal R(\phi)=\Delta I(\phi)=-\mathcal
R(-\phi)$ and $\mathcal D(\phi)=\mathcal D(-\phi)$. In other words,
we show that odd (even) conductance terms are even (odd) in the
magnetic flux. 
 Note that ``spatial" or ``geometrical" symmetry
refers here to the left-right mirror symmetry of the junction.
Below we refer to these symmetries as the ``MF symmetry relations" .

(iii) {\it Magnetic field-Gate voltage (MFGV) symmetry relations beyond linear response.}
In Secs. \ref{violation}-\ref{simulation} we focus on
geometrically {\it asymmetric} setups, adopting the double dot AB
interferometer as an example.
While we demonstrate, using numerical
simulations, the breakdown of the MF relations 
under spatial asymmetry, in Appendix A
we prove that charge conjugation symmetry entails
magnetic field-gate voltage symmetries: 
$\mathcal R(\epsilon_d,\phi)=-\mathcal R(-\epsilon_d,-\phi)$,
and $\mathcal D(-\epsilon_d,-\phi)=\mathcal D(\epsilon_d,\phi)$,
with $\epsilon_d$ as the double-dot energies.
We refer below to these symmetries as the ``MFGV symmetry relations" .


(iv) In Appendix B we prove that the heat current (within a
heat-conserving setup) satisfies relations
analogous to (i)-(iii).


\section{Phase rigidity and absence of rectification}
\label{dephasing}

The Onsager-Casimir symmetry $I(\phi)=I(-\phi)$ is
preserved  under dephasing effects, implemented via
a dephasing probe, even beyond the linear response regime. It is also satisfied in
the presence of elastic and inelastic effects, implemented using the voltage
probe technique, only as long the system is maintained in the linear
response regime. These results have already been discussed in
e.g., Ref. \cite{Buttiker-4}. We details these steps here
so as to provide closed expressions for the probe parameters in the linear response regime.


\subsection{Dephasing effects beyond linear response}
\label{rigidity1}

Implementing the dephasing probe (\ref{eq:Dprobe}) we obtain the
respective distribution
\bea
f_P(\phi)=\frac{\mathcal T_{L,P}f_L +  \mathcal T_{R,P}f_R   }{
\mathcal T_{P,L} + \mathcal T_{P,R}}.
\label{eq:dephP}
\eea
We substitute this function into Eq. (\ref{eq:DeltaI}), a measure
for phase asymmetry, and obtain
\bea
\Delta I&=& \frac{1}{2}
\int  \left[ \mathcal T_{L,R} - \mathcal T_{R,L}\right] f_R d\epsilon
\nonumber\\
&+& \frac{1}{2}
\int \mathcal T_{L,P}
\frac{ \mathcal T_{P,L} f_L + \mathcal T_{P,R}f_R }{ \mathcal T_{L,P} +
\mathcal T_{R,P}} d\epsilon
\nonumber\\
&-& \frac{1}{2}
\int  \mathcal T_{P,L}
\frac{ \mathcal T_{L,P} f_L + \mathcal T_{R,P}f_R }{ \mathcal T_{P,L} +
\mathcal T_{P,R}} d\epsilon
\eea
The denominators in these integrals are identical, see Eq.
(\ref{eq:R2}), thus  we combine the last two terms
\bea
\Delta I&=&\frac{1}{2}
\int  \left[ \mathcal T_{L,R} - \mathcal T_{R,L}\right] f_R d\epsilon
\nonumber\\
&+&\frac{1}{2}
\int
\frac{ \left[ \mathcal T_{L,P} \mathcal T_{P,R} - \mathcal T_{P,L} \mathcal T_{R,P} \right]f_R  }
{ \mathcal T_{P,R} +
\mathcal T_{P,L}} d\epsilon
\eea
Utilizing Eq. (\ref{eq:R2}) in the form $\mathcal T_{L,P}$ =
$\mathcal T_{P,L} + \mathcal T_{P,R}-\mathcal T_{R,P}$, we organize
the numerator of the second integral, $(\mathcal T_{P,R}  - \mathcal
T_{R,P} )   (\mathcal T_{P,R}  + \mathcal T_{P,L} ) f_R $. This
results in
\bea
\Delta I&=&\frac{1}{2}
\int  \left[ \mathcal T_{L,R} - \mathcal T_{R,L}
+
\mathcal T_{P,R}  - \mathcal T_{R,P} \right] f_R d\epsilon
\nonumber\\
&=&\frac{1}{2}
\int  f_R \left[\sum_{\nu\neq R}\mathcal T_{\nu,R} -
\sum_{\nu\neq R}\mathcal T_{R,\nu}\right] d\epsilon,
\eea
which is identically zero, given Eq. (\ref{eq:R2}).
This concludes our proof that dephasing effects,
implemented via a dephasing probe, cannot break phase rigidity
even in the nonlinear regime.

Following similar steps we show that elastic dephasing effects
cannot generate the effect of charge rectification even when the
junction acquires spatial asymmetries. We substitute $f_P$
into Eq. (\ref{eq:IL}) and obtain
\bea
I_L=\int[ F_L f_L - F_R f_R] d\epsilon
\eea
with
\bea
F_L=\frac{\mathcal T_{L,R}(\mathcal T_{P,L}+\mathcal T_{P,R})
+ \mathcal T_{L,P}\mathcal T_{P,R}}{(\mathcal T_{P,L}+\mathcal T_{P,R})}.
\eea
$F_R$ is defined analogously, interchanging  $L$ by $R$.
We now note the following identities,
\bea&&
\mathcal T_{L,P}\mathcal T_{P,R}=
[\mathcal T_{P,L}+ \mathcal T_{P,R}- \mathcal T_{R,P}] \mathcal T_{P,R}
\nonumber\\
&&=
(\mathcal T_{P,R}- \mathcal T_{R,P})(\mathcal T_{P,R}+ \mathcal T_{P,L})
+\mathcal T_{P,L} \mathcal T_{R,P}
\nonumber\\
&&=
(\mathcal T_{R,L}-\mathcal T_{L,R})
(\mathcal T_{P,R}+ \mathcal T_{P,L})
+\mathcal T_{P,L} \mathcal T_{R,P}
\eea
Reorganizing the first and third lines we find that
\bea &&\mathcal T_{L,R}(\mathcal T_{P,R}+ \mathcal T_{P,L})
+ \mathcal T_{L,P}\mathcal T_{P,R}
\nonumber\\
&&=
\mathcal T_{R,L} (\mathcal T_{P,R}+ \mathcal T_{P,L})+
\mathcal T_{P,L} \mathcal T_{R,P}
\eea
which immediately implies that $F_L=F_R$. This is turn
leads to $I=-\bar I$, thus $\mathcal R=0$. We
conclude that the current only includes odd (linear and nonlinear) conductance terms
under  elastic dephasing,
$I= \mathcal D(\phi)=\mathcal D(-\phi)$, and that phase rigidity is maintained
even if spatial asymmetry is
presented. This conclusion in valid under either a voltage or a temperature bias. 

\subsection{Inelastic effects in linear response}
\label{rigidity2}

We introduce inelastic effects using the voltage probe technique. In
the linear response regime we expand the $\nu=L,R,P$ Fermi functions
around the equilibrium state
$f_a(\epsilon)=[e^{\beta_a(\epsilon-\mu_a)}+1]^{-1}$,
\bea f_{\nu}(\epsilon)=f_a(\epsilon)
-(\mu_{\nu}-\mu_a)\frac{\partial f_a}{\partial \epsilon}. \eea
The three terminals are maintained at the same temperature $T_a$.
The derivative $\frac{\partial f_a}{\partial \epsilon}$ is evaluated
at the equilibrium value $\mu_a$. For simplicity we set $\mu_a=0$.
We enforce the voltage probe condition, $I_P=0$, demanding that
\bea
\int \left[
\left(\mathcal T_{P,L}+\mathcal T_{P,R}\right ) f_P(\phi)
- \mathcal T_{L,P}f_L -\mathcal T_{R,P}f_R \right]d\epsilon=0.
\label{eq:VoltageP}
\eea
In linear response this translates to
\bea
0&=& \int  \Big[ \left(\mathcal T_{P,L}+\mathcal T_{P,R}\right )
\left( f_a-\mu_P(\phi) \frac{\partial f_a}{\partial \epsilon} \right)
\nonumber\\
&-& \mathcal T_{L,P} \left( f_a-\mu_L \frac{\partial f_a}{\partial \epsilon} \right)
-\mathcal T_{R,P} \left( f_a-\mu_R \frac{\partial f_a}{\partial \epsilon} \right)
\Big]d\epsilon.
\nonumber\\
\label{eq:probeLR}
\eea
For convenience, we apply the voltage symmetrically,
$\mu_L=-\mu_R=\Delta \mu/2$. We organize Eq. (\ref{eq:probeLR}) and
obtain the probe chemical potential, a linear function in $\Delta
\mu$,
\bea
\mu_P(\phi)= \frac{\Delta \mu}{2}  \frac{\int d\epsilon \frac{\partial f_a}{\partial \epsilon}
\left(\mathcal T_{L,P}-\mathcal T_{R,P}\right )  }
{\int d\epsilon \frac{\partial f_a}{\partial \epsilon} \left(\mathcal T_{P,L}+\mathcal T_{P,R}\right ) }.
\label{eq:muP}
\eea
We simplify this result by introducing a short notation for the conductance between the $\nu$
and $\xi$ terminals,
\bea G_{\nu,\xi}(\phi) \equiv \int d\epsilon \left(-\frac{\partial
f_a}{\partial\epsilon}\right) \mathcal T_{\nu,\xi}(\epsilon,\phi).
\label{eq:G} \eea
This quantity fulfills relations analogous to Eqs.
(\ref{eq:R1}) and (\ref{eq:R2}) \cite{Buttiker}. For brevity, we do not write next
the phase variable in $G$, evaluating it at the phase
$\phi$ unless otherwise mentioned.
The probe potential can now be compacted, 
\bea \mu_P(\phi)= \frac{\Delta \mu}{2} \frac{G_{L,P} -
G_{R,P}}{G_{P,L}+G_{P,R}}. 
\label{eq:muPGG}
\eea
Furthermore, in geometrically symmetric systems $\mathcal
T_{R,P}(\epsilon, \phi)=\mathcal T_{P,L}(\epsilon,\phi)$, resulting in
$G_{R,P}(\phi)=G_{P,L}(\phi)$ and
\bea \mu_P(\phi)=
\frac{\Delta \mu}{2} \frac{
G_{L,P}(\phi)-G_{L,P}(-\phi)  } {G_{L,P}(\phi)+G_{L,P}(-\phi) }.
\eea
Thus
$\mu_P(\phi)=-\mu_P(-\phi)$ in linear response. Below we show that
this symmetry does not hold far-from-equilibrium. We now expand Eq.
(\ref{eq:DeltaI}) in the linear response regime
\bea
\Delta I&=&\frac{1}{2} \int
\left( \mathcal T_{L,R}-\mathcal T_{R,L} \right)
\left( f_a-\mu_R \frac{\partial f_a}{\partial \epsilon}
\right)d\epsilon
\nonumber\\
&+& \frac{1}{2}
\int  \Bigg\{
\mathcal T_{L,P} \left[f_a - \mu_P(-\phi) \frac{\partial f_a}{\partial \epsilon} \right]
\nonumber\\
&-& \mathcal T_{P,L} \left[f_a - \mu_P(\phi) \frac{\partial f_a}{\partial \epsilon} \right]
\Bigg \} d\epsilon.
\eea
Utilizing the definition (\ref{eq:G}) we 
compact this expression, 
\bea \Delta I&=&\frac{1}{2}\left(G_{L,R}-G_{R,L} \right)(\mu_R)
\nonumber\\
&-& \frac{1}{2} \left[ G_{P,L}\mu_P(\phi) -G_{L,P}\mu_P(-\phi)
\right]. \label{eq:V} \eea
Using Eq. (\ref{eq:R2}), the first line can be rewritten as
\bea I_1= \frac{\Delta \mu}{4}\left(  G_{L,P}-G_{P,L}   \right).
\eea
The second line in Eq. (\ref{eq:V}) reduces to
\bea I_2&=& -\frac{\Delta \mu}{4} G_{P,L}
\frac{G_{L,P}-G_{R,P}}{\mathcal N} +\frac{\Delta \mu}{4} G_{L,P}
\frac{G_{P,L}-G_{P,R}}{\mathcal N}
\nonumber\\
&=& -\frac{\Delta \mu}{4}  \frac{G_{L,P} G_{P,R}-G_{P,L}
G_{R,P}}{\mathcal N} \eea
where we have introduced the short notation $\mathcal N \equiv
G_{P,L}+G_{P,R}$.
Now, we substitute $G_{L,P}= G_{P,L} +
G_{P,R} - G_{R,P}$, and this allows us to write
\bea I_2&=& -\frac{\Delta \mu}{4} \frac{(G_{P,R}- G_{R,P})(G_{P,R}
+G_{P,L})}{\mathcal N}
\nonumber\\
&=& -\frac{\Delta \mu}{4} (G_{P,R}- G_{R,P}) \eea
%
Combining $\Delta I=I_1+I_2$, we reach
\bea \Delta I &=& -\frac{\Delta \mu}{4} \left(G_{P,R}- G_{R,P}
-G_{L,P}+G_{P,L}\right)
\nonumber\\
&=& -\frac{\Delta \mu}{4} \left( \sum_{\nu\neq
P}G_{P,\nu}-\sum_{\nu\neq P}G_{\nu,P} \right)
\eea
which is identically zero given the conductance conservation
(\ref{eq:R2}). It is trivial to note that no rectification
takes place in the linear response regime, $\mathcal R=0$.

\section{Beyond linear response:  spatially symmetric setups}
\label{symmetry}

In this section we consider the role of inelastic effects on the
current symmetry in an AB interferometer, beyond the linear response
regime. The probe condition $I_P=0$
translates  Eq. (\ref{eq:IP}) into  three relations,
\bea\int d\epsilon (\mathcal T_{P,L}+\mathcal T_{P,R})f_ P(\phi)= \int d\epsilon (\mathcal T_{L,P}f_L +\mathcal
T_{R,P}f_R)
 \nonumber\\
\int d\epsilon(\mathcal T_{L,P}+\mathcal T_{R,P}) f_P(-\phi)= \int d\epsilon( \mathcal T_{P,L}f_L +\mathcal
T_{P,R}f_R)
\nonumber\\
\int d\epsilon(\mathcal T_{P,L}+\mathcal T_{P,R})\bar f_P(\phi)=\int d\epsilon\left(\mathcal T_{L,P}f_R+\mathcal
T_{R,P}f_L\right).
\label{eq:fpp}
\eea
%
%
First, we consider the situation when time reversal symmetry is protected
with the magnetic phase given by multiples of
$2\pi$. Then, $\mathcal T_{\nu,\xi}= \mathcal T_{\xi,\nu}$, and particularly we note that
$\mathcal T_{L,P}=\mathcal T_{P,L}$. Furthermore, in the  model
considered in Sec. \ref{violation}, $\mathcal T_{P,L}=\chi \mathcal
T_{P,R}$, with $\chi$ as an energy independent parameter,
reflecting spatial asymmetry,
see for example the discussion around Eq.
(\ref{eq:gamma}). Using the voltage probe condition (\ref{eq:fpp})
we find that
\bea
&&\int
 (\mathcal T_{P,L}+\mathcal T_{P,R})(f_P(\phi)+\bar f_P(\phi))d\epsilon
\nonumber\\
&&= \int (\mathcal T_{P,L}+\mathcal T_{P,R})(f_L+f_R)d\epsilon,
\eea
Given the linear relation between $\mathcal T_{L,P}$ and $\mathcal
T_{R,P}$, this equality holds separately for each transmission
function,
\bea && \int \mathcal T_{P,\nu}(f_P(\phi)+\bar f_P(\phi))d\epsilon
\nonumber\\
&&= \int \mathcal T_{P,\nu}(f_L+f_R)d\epsilon \,\,\,\,\, \nu=L,R,
\label{eq:RRR0}
\eea
providing $\mathcal R=0$ in Eq. (\ref{eq:R}). Thus,  if $\mathcal T_{P,L}= \mathcal T_{L,P}=
\chi \mathcal T_{P,R}$,
rectification is absent. 
In physical terms,  the junction conducts symmetrically for forward and reversed direction,
though many-body effects are presented,
if we satisfy two conditions: (i) Spatial asymmetry is included in an energy-independent manner,
for example using different broad-band hybridization parameters at the two ends.
(ii) Time reversal symmetry is protected.

We now derive symmetry relations for left-right symmetric systems with broken time-reversal symmetry.
In this case the
mirror symmetry $\mathcal T_{P,L}(\phi)=\mathcal T_{P,R}(-\phi)$
applies, translating to
\bea
\mathcal T_{P,L}(\phi)=\mathcal T_{R,P}(\phi).
\label{eq:R3}
\eea
When used in  Eq. (\ref{eq:fpp}), we note that 
the distributions should obey
\bea
\bar f_P(\phi)=f_P(-\phi),
\label{eq:fpS}
\eea
leading to $\bar \mu_P(\phi)=\mu_P(-\phi)$. We emphasize that
$\mu_P(\phi)$ itself does not posses a phase symmetry.

Since charge dissipation is not allowed,
the deviation from phase rigidity, Eq. (\ref{eq:DeltaI}), can
be also expressed in terms of the current $I_R$, to provide (note the sign convention)
\bea
\Delta I(\phi)&=& \frac{1}{2}\int d\epsilon[(\mathcal T_{L,R}-\mathcal T_{R,L})f_L
\nonumber\\
&-&\mathcal T_{R,P}f_P(-\phi) +\mathcal T_{P,R}f_P(\phi)].
\label{eq:DeltaI2}
\eea
We define $\Delta I$ by the average of Eqs.
(\ref{eq:DeltaI}) and (\ref{eq:DeltaI2}),
\bea \Delta I(\phi)&=&\frac{1}{4}\int d\epsilon\Big[(\mathcal
T_{L,R}-\mathcal T_{R,L})(f_L+f_R)
\nonumber\\
&+&(\mathcal T_{L,P}-\mathcal T_{R,P})f_P(-\phi) +(\mathcal
T_{P,R}-\mathcal T_{P,L})f_P(\phi)\Big]. \nonumber \eea
We proceed and make use of two relations: $\mathcal
T_{L,R}-\mathcal T_{R,L} = \mathcal T_{P,L}-\mathcal T_{L,P}$.
and Eq (\ref{eq:R3}), valid in geometrically
symmetric junctions. With this at hand we write
\bea
&& \Delta I(\phi)= \frac{1}{4}\int (\mathcal T_{P,L}-\mathcal
T_{P,R})(f_L+f_R-f_P(\phi)-\bar f_P(\phi)) d\epsilon
\nonumber\\
&&= \mathcal R(\phi)=- \mathcal R(-\phi),
\eea
%
This concludes our derivation that
\bea
\Delta I(\phi)=\mathcal R(\phi)=-\mathcal R(-\phi),\,\,\,\,\,
\mathcal D(\phi)=\mathcal D(-\phi).
\label{eq:DI}
\eea
%
In spatially
symmetric systems odd conductance terms acquire even symmetry with
respect to the magnetic field,
as noted experimentally \cite{breakE2, breakE4},
while even conductance terms, constructing $\mathcal R$, are odd
with respect to $\phi$.
The relation $\Delta I(\phi)=\mathcal R(\phi)$
could be exploited in experimental studies: One could
determine whether a quantum dot junction is $L$-$R$
symmetric by testing this equality.

The following observations can be made:
First, Eq. (\ref{eq:DI}) does not hold when a spatial asymmetry is introduced, by
coupling the scattering centers unevenly to the leads. Second,
the symmetry relations obtained here are valid
under the more restrictive (non-dissipative) voltage-temperature
probe, Eq. (\ref{eq:TVprobe}). Finally, the  
analysis in this section reveals sufficient conditions for charge rectification
for structurally {\it symmetric} junctions:
$\mathcal R(\phi)\neq 0$  when time-reversal symmetry is broken, $\phi\neq
2\pi n$, and inelastic scatterings are allowed.



\section{Beyond linear response: spatially asymmetric setups}

\label{violation}

\begin{figure}[htpb]
\vspace{-12mm} \hspace{6mm}
{\hbox{\epsfxsize=180mm\epsffile{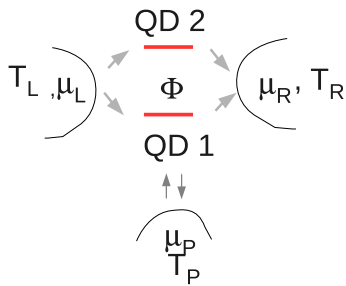}}}
\vspace{-90mm}
\caption{
Scheme of a double-dot AB interferometer
susceptible to many-body effects.
The two quantum dots (QD) are each represented by a single electronic level,
which do not directly couple. The total magnetic flux is denoted by
$\Phi$. Dot  '1' may be  susceptible to elastic dephasing and inelastic
effects, introduced here through the coupling of this dot to a probe, the
terminal $P$. 
}
\label{scheme2d}
\end{figure}

We adopt a double-dot AB model, see Fig. \ref{scheme2d},
allow for inelastic effects and spatial asymmetry, and 
prove analytically the validity of magnetic field - gate voltage symmetries.


\subsection{Model: Double-dot Interferometer}
\label{model}

We focus on an AB setup with a quantum
dot located at each arm of the interferometer. The dots are
connected to two metal leads maintained in a biased state. For
simplicity, we do not consider electron-electron
interactions and the Zeeman effect, thus, we can ignore the
spin degree of freedom and describe each quantum dot by a
spinless electronic level. The total Hamiltonian includes the
following terms,
\bea
H=H_S+\sum_{\nu=L,R,P}H_{\nu} +
\sum_{\nu=L,R} H_{S, \nu}
+H_{S,P}
\label{eq:H}
\eea
The subsystem Hamiltonian includes two (uncoupled) electronic
states
\bea H_S=\sum_{n=1,2}\epsilon_na_n^\dagger a_n.
\eea
The three reservoirs (metals) comprise of a
collection of non-interacting electrons,
\bea 
H_{\nu}=\sum_{j\in \nu}\epsilon_ja_j^{\dagger}a_j.
\eea

Here $a_{j}^{\dagger}$ ($a_{j}$)  are fermionic creation
(annihilation) operators of electrons with momentum $j$ and energy
$\epsilon_j$. $a_{n}^{\dagger}$ and $a_{n}$ are the respective
operators for the dots.
The subsystem-bath coupling terms are given by
\beq H_{S,L}+H_{S,R}= \sum_{n,l} v_{n,l} a_{n}^{\dagger}a_{l}
e^{i\phi_{n}^{L}}+\sum_{n,r}v_{n,r}a_{r}^{\dagger}a_{n}
e^{i\phi_{n}^{R}}+h.c..
\nonumber
\eeq
We assume that only dot '1' couples to the probe
\bea
H_{S,P}= \sum_{p} v_{1,p} a_{1}^{\dagger}a_{p}+ h.c.
\label{eq:HSP}
\eea
Here $v_{n,j}$ is the coupling strength of dot $n$ to the $j$ state
of the $J$ bath. Below we assume that this parameter does not
depend on the $n=1,2$ level index. $\phi_{n}^{L}$ and $\phi_{n}^{R}$
are the AB phase factors, acquired by electron waves in a magnetic
field perpendicular to the device plane. These phases are
constrained to satisfy
\beq
\phi_{1}^{L}-\phi_{2}^{L}+\phi_{1}^{R}-\phi_{2}^{R}=\phi 
\eeq
%
and we adopt the gauge
$\phi_{1}^{L}-\phi_{2}^{L}=\phi_{1}^{R}-\phi_{2}^{R}=\phi/2$.

We voltage-bias the system, $\Delta
\mu\equiv\mu_L-\mu_R$, with $\mu_{L,R}$ as the chemical
potential of the metals, and use the convention that a positive current is flowing left-to-right.
While we bias the system in a
symmetric manner, $\mu_L=-\mu_R$, this choice does not limit the generality of
our discussion since the dots  may be
gated away from the so called ``symmetric point" at which
$\mu_L-\epsilon_{n}=\epsilon_{n}-\mu_R$.


\subsection{Green's function expressions}
\label{GreenF}

Our model does not include interacting particles, thus its
steady-state characteristics can be written exactly using the
nonequilibrium Green's function approach \cite{Wingreen,RevGreen}.
Transient effects were recently explored in Refs. \cite{Salil1,
Ora-time}. Since relevant derivations were detailed in our recent 
study \cite{Salil2}, we only include here the principal expressions.
In terms of the Green's function, the transmission coefficient is
defined as
\bea
\mathcal T_{\nu,\xi}={\rm
Tr}[\Gamma^{\nu}G^+\Gamma^{\xi}G^-],
\eea
where the trace is performed over the states of the subsystem (dots).
In our model the matrix $G^{+}$ ($G^-=[G^+]^{\dagger}$) takes the
form
\bea G^{+}=\left[  \begin{array}{cc}
 \epsilon-\epsilon_{1}+\frac{i(\gamma_L+\gamma_R+\gamma_P)}{2} & \frac{i\gamma_L}{2}e^{i\phi/2}
+\frac{i\gamma_R}{2}e^{-i\phi/2}\\
\frac{i\gamma_L}{2}e^{-i\phi/2}+\frac{i\gamma_R}{2}e^{i\phi/2}
&  \epsilon-\epsilon_{2}+\frac{i(\gamma_L+\gamma_R)}{2}\\
 \end{array}\right]^{-1},
\nonumber \eea
with the hybridization matrices satisfying
\bea &&\Gamma^{L}=\gamma_L\left[ \begin{array}{cc}
              1 & e^{i\phi/2}\\
              e^{-i\phi/2} &1\\
\end{array}\right],\,\,\,\,
\Gamma^{R}=\gamma_R \left[ \begin{array}{cc}
              1 & e^{-i\phi/2}\\
              e^{i\phi/2} &1\\
\end{array}\right]
\nonumber\\
&& \Gamma^{P}=\gamma_P\left[ \begin{array}{cc}
              1 & 0\\
              0 & 0\\
\end{array}\right]
\label{eq:gamma}
\eea
The coupling energy between the dots and leads is given by
$\gamma_{\nu}(\epsilon)=2\pi\sum_{j\in\nu}|v_j|^2\delta(\epsilon-\epsilon_j)$.
In the wide-band limit adopted in this work
$\gamma_{\nu}$ are taken as energy independent parameters.


We now assume that the dots are energy-degenerate,
$\epsilon_d\equiv\epsilon_1=\epsilon_2$,
but allow for spatial asymmetry in the form $\gamma_L\neq\gamma_R$.
The transmission functions follow a simple form,
\bea
&&\mathcal{T}_{LR}=
\frac{\gamma_{L}\gamma_{R}}{\Delta(\epsilon,\phi)}\left[4(\epsilon-\epsilon_{d})^2\cos^2\frac{\phi}{2}+\frac{\gamma_{P}^2}{4}+
\gamma_{P}(\epsilon_{d}-\epsilon)\sin{\phi}\right] \nonumber\\
&&\mathcal{T}_{LP}=
\frac{\gamma_{L}\gamma_{P}}{\Delta(\epsilon,\phi)}\left[(\epsilon-\epsilon_{d})^2
+{\gamma_{R}^2}\sin^2{\frac{\phi}{2}}+\gamma_{R}(\epsilon-\epsilon_{d})\sin{\phi}\right]  \nonumber\\
&&\mathcal{T}_{RP}=
\frac{\gamma_{R}\gamma_{P}}{\Delta(\epsilon,\phi)}\left[(\epsilon-\epsilon_{d})^2
+{\gamma_{L}^2}\sin^2{\frac{\phi}{2}}
-\gamma_{L}(\epsilon-\epsilon_{d})\sin{\phi}\right]
\nonumber\\
  \label{eq:TRP}
\eea
with the denominator an even function of $\phi$,
\bea
\Delta(\epsilon,\phi)&=&
\left[(\epsilon-\epsilon_d)^2-\gamma_L\gamma_R\sin^2{\frac{\phi}{2}}-
\frac{(\gamma_L+\gamma_R)\gamma_P}{4}\right]^2
\nonumber\\
&+&\left(\gamma_L+\gamma_R + \frac{\gamma_P}{2}\right)^2 (\epsilon-\epsilon_d)^2.
\eea
\label{eq:Deno}
It is trivial to confirm that
in the absence of the probe the even symmetry of the current with $\phi$
is satisfied, beyond linear response \cite{Kubala}.
\bea
&&I_L(\phi)=
\nonumber\\
&&\int
d\epsilon\frac{4\gamma_L\gamma_R(\epsilon-\epsilon_d)^2\cos^2\frac{\phi}{2} [f_L(\epsilon)-f_R(\epsilon)]}
{\left[(\epsilon-\epsilon_d)^2-\gamma_L\gamma_R\sin^2\frac{\phi}{2}\right]^2+(\gamma_L+\gamma_R)^2(\epsilon-\epsilon_d)^2}
\nonumber
\eea
%
%
%
%
With the probe, inspecting the transmission functions in conjunction with Eq. (\ref{eq:muPGG}),
we immediately conclude that under spatial asymmetries
the probe chemical potential does not obey a particular symmetry,
even in linear response 
when phase rigidity is trivially obeyed, see Sec. \ref{rigidity2}.

We now discuss the properties of the probe when the interferometer is $L$-$R$ symmetric,
$\gamma/2=\gamma_L=\gamma_R$. The transmission functions satisfy
$\mathcal T_{R,P}(\epsilon,\phi) = \mathcal T_{P,L}(\epsilon,\phi)$. 
%
%
We substitute these expressions into Eq. (\ref{eq:dephP}) and resolve
the dephasing probe distribution \cite{Salil2}
\bea
 f_P^D(\epsilon,\phi)&=& \frac{f_L(\epsilon)+f_R(\epsilon)}{2}
\nonumber\\
& +& \frac{\gamma(\epsilon-\epsilon_d)\sin \phi
}   {4\Big[(\epsilon-\epsilon_d)^2 + \omega_0^2\Big]}
[f_L(\epsilon)-f_R(\epsilon)]
\label{eq:fpDD} \eea
with $\omega_0=\frac{\gamma}{2}\sin \frac{\phi}{2}$.
The nonequilibrium term in this distribution is {\it odd} in the magnetic
flux. Similarly, when a voltage probe ($V$) is implemented, analytic results can be obtained
in the linear response regime,
%
\bea
\mu_P^V(\phi)= \Delta \mu  \sin \phi
\frac{\int d\epsilon \frac{\partial f_a}{\partial \epsilon}
\frac{\gamma(\epsilon-\epsilon_d)}{\Delta(\epsilon,\phi)}}
{\int d\epsilon \frac{\partial f_a}{\partial \epsilon}
\frac{2(\epsilon-\epsilon_d)^2 +\frac{1}{2}\gamma^2\sin^2 \frac{\phi}{2}}{\Delta(\epsilon,\phi)}}.
\label{eq:muPG}
\eea
Here $f_a$ stands for the equilibrium (zero bias) Fermi-Dirac
function. This chemical potential is an {\it odd} function of
the magnetic flux, though phase rigidity is maintained in the
linear response regime.

%
%
%
%
%

\subsection{Generalized magnetic field-gate voltage symmetries}
\label{holeS}

The MF symmetry relations (\ref{eq:DI}) are not obeyed when the spatial mirror symmetry is broken.
Instead, in Appendix A we prove that in a generic model for a  double-dot interferometer 
susceptible to inelastic effects
the following result holds
\bea
\mathcal R=-\mathcal C (\mathcal R), \,\,\,\, \mathcal D=\mathcal C (\mathcal D).
\eea
Here $\mathcal C$ stands for the charge conjugation operator, transforming electrons to holes 
and vise versa.
In terms of the interferometer parameters this relation reduces to the 
following magnetic flux-gate voltage (MFGV) symmetries, 
\bea
\mathcal R(\epsilon_d,\phi)&=&-\mathcal R(-\epsilon_d,-\phi),
\nonumber\\
\mathcal D(\epsilon_d,\phi)&=&\mathcal D(-\epsilon_d,-\phi).
\label{eq:CC}
\eea
Since the energies of the dots can be modulated with a gate voltage \cite{Hernandez11},
these generalized symmetries can be examined experimentally.

\section{Numerical simulations}
\label{simulation}
Using numerical simulations we demonstrate the behavior of 
the voltage and the voltage-temperature probes far-from-equilibrium, and the implications on 
phase rigidity and magnetic field symmetries.

\subsection{Probe parameters}
\label{simulation-probe}

We consider the model Eq. (\ref{eq:H}) and implement inelastic
effects with the dissipative voltage probe, by solving the probe
condition (\ref{eq:Vprobe}) numerically-iteratively, using Eq.
(\ref{eq:NR}), to obtain $\mu_P$. We also investigate the transport behavior of
the model under the
more restrictive dissipationless voltage-temperature probe, by
solving Eq. (\ref{eq:TVprobes}) to obtain both $\mu_P$ and $T_P$.

\begin{figure}[htbp]
\hspace{2mm} 
{\hbox{\epsfxsize=80mm \epsffile{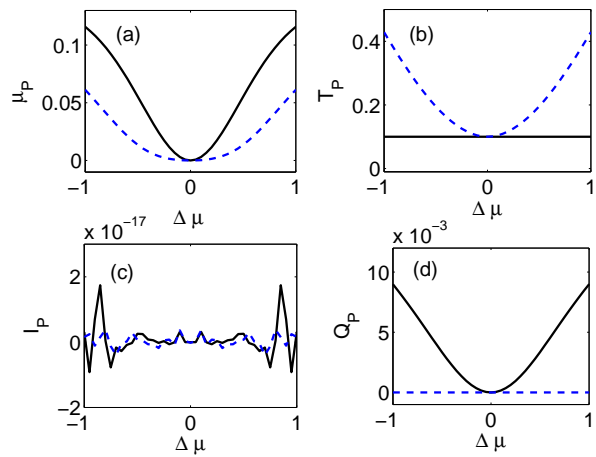}}} \caption{
Self-consistent parameters of the voltage probe (full) and the
voltage-temperature probe (dashed), displaying disparate behavior
far-from-equilibrium: (a) Probe chemical potential, (b) temperature.
We also show (c) the magnitude of net charge current from the probe and
(d) net heat current from the conductor towards the probe. The interferometer consists
two degenerate levels with $\epsilon_{1,2}=0.15$ coupled evenly to
the metal leads $\gamma_{L,R}=0.05$. Other parameters are
$\gamma_P=0.1$, $\phi=0$, and $T_L=T_R=0.1$. The probe temperature
is set at $T_P=0.1$ in the voltage probe case.} \label{probe1}
\end{figure}

Fig. \ref{probe1} displays the self-consistent probe parameters $\mu_P$
and $T_P$ for $\phi=0$ when heat dissipation is allowed at the probe (full line),
and when neither heat nor charge dissipation takes place within $P$ (dashed line). We find
that the probe parameters largely vary depending on the probe
condition, particularly at high biases when significant heat
dissipation can take place [panel (d)]. We also
verify that when Newton-Raphson iterations converge, the charge current to the
probe is negligible, $|I_P/I_L|<10^{-12}$. Similarly, the heat
current in the voltage-temperature probe is negligible once
convergence is reached.

\begin{figure}[htbp]
\hspace{2mm} 
{\hbox{\epsfxsize=80mm \epsffile{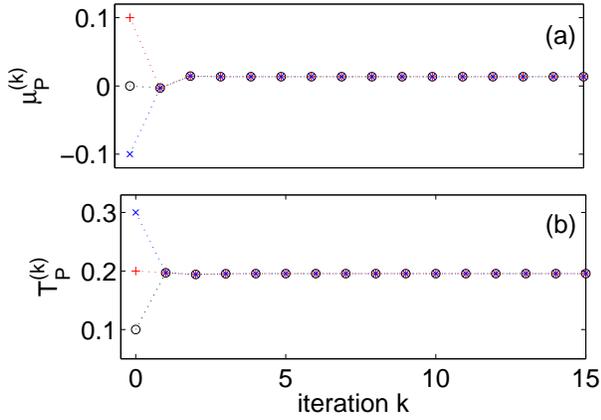}}}
\caption{
Insensitivity of the parameters of the voltage-temperature probe [Eq. (\ref{eq:TVprobes})]
to initial conditions. 
(a) chemical potential of the probe and (b) its temperature.
The different initial conditions are identified by the values at the first iteration.
The interferometer's parameters follow Fig. \ref{probe1} with $\Delta \mu=0.5$
and $T_L=T_R=0.1$.
}
\label{probe2}
\end{figure}

Uniqueness of the parameters of the voltage
and temperature probes has been recently proved in Ref.
\cite{Unique}. We complement this analytical analysis and
demonstrate that the parameters of the voltage-temperature probe 
are insensitive to the initial conditions
adopted, see Fig. \ref{probe2}.
Convergence has been typically achieved with $\sim 5$
iterations. While the voltage probe had
easily converged even at large biases, we could  not manage to
converge the voltage-temperature probe parameters at large biases
$\Delta \mu>1$ and low temperatures $T_{L,R}<1/50$ since eliminating heat
dissipation within the probe requires extreme values,
leading to numerical divergences within the model parameters adopted.


\subsection{Nonlinear transport with dissipative inelastic effects}
\label{simulation-transport}

\begin{figure}[htbp]
\hspace{2mm} 
{\hbox{\epsfxsize=80mm \epsffile{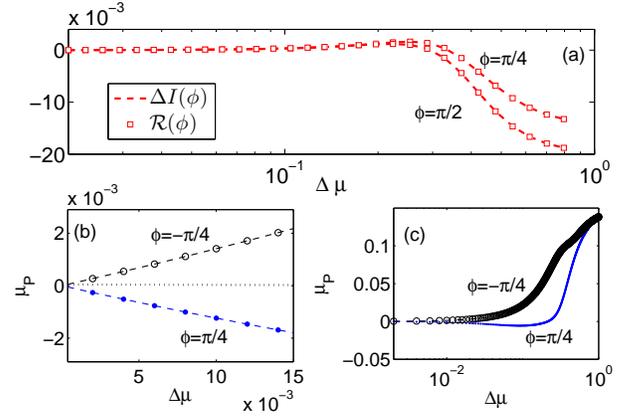}}} \caption{(a) MF
symmetry and rectification in spatially symmetric junctions.
(b) Probe chemical potential in the linear response regime. (c)
Probe chemical potential beyond linear response. The junction's
parameters are $\epsilon_1=\epsilon_2=0.15$, $\gamma_P=0.1$,
$\beta_a=50$ and $\gamma_L=\gamma_R=0.05$.} \label{Fig1}
\end{figure}

In this subsection we examine 
the nonlinear transport behavior of an AB interferometer coupled to a voltage probe.
In Fig. \ref{Fig1} (a) we display the measure
$\Delta I$ as a function of bias for a spatially symmetric
system using two representative phases, $\phi=\pi/2$ and $\pi/4$. We
confirm numerically that in the linear response regime $\Delta I=
0$. More generally, the relation $\Delta I=\mathcal R$
is satisfied for all biases, as expected from Eq. (\ref{eq:DI}). Our
conclusions are intact when an ``up-down" asymmetry is implemented
in the form $\epsilon_1\neq\epsilon_2$ \cite{Salil3}. The
corresponding chemical potential of the probe is shown in Fig.
\ref{Fig1}(b)-(c) for $\phi=\pi/4$. In the linear response regime
it grows linearly with $\Delta \mu$ and it obeys an odd symmetry
relation, $\mu_P(\phi)=-\mu_P(-\phi)$. Beyond linear response
$\mu_P$ does not follow neither an even nor an odd phase symmetry,
but at large enough biases it is independent of the sign of the phase.

\begin{figure}[htbp]
\hspace{2mm} {\hbox{\epsfxsize=80mm \epsffile{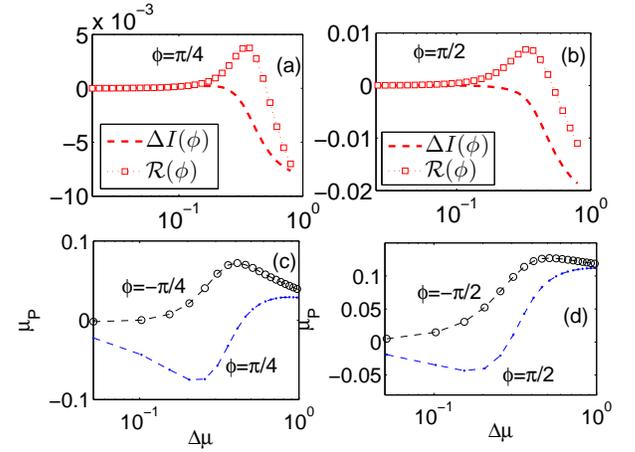}}} \caption{
Breakdown of the MF symmetry relations for spatially asymmetric junctions,
$\gamma_L=0.05\neq\gamma_R=0.2$. (a)-(b)
$\Delta I$ (dashed) and $\mathcal R$ (square) for $\phi=\pi/4$ and $\pi/2$.
%
The corresponding probe potential is displayed in panel (c) for
$\phi=\pm\pi/4$ and in panel (d) for $\phi=\pm\pi/2$. Other
parameters are $\epsilon_1=\epsilon_2=0.15$, $\gamma_P=0.1$ and
$\beta_a=50$. } \label{Fig2}
\end{figure}

Fig. \ref{Fig2} displays results when spatial asymmetry in the form $\gamma_L\neq \gamma_R$
is implemented.
Here we observe that $\Delta I\neq \mathcal R$, and that
the probe chemical potential does not satisfy an odd symmetry with
the magnetic phase, even in the linear response regime. 

\begin{figure}[t]
\hspace{1mm} 
{\hbox{\epsfxsize=80mm \epsffile{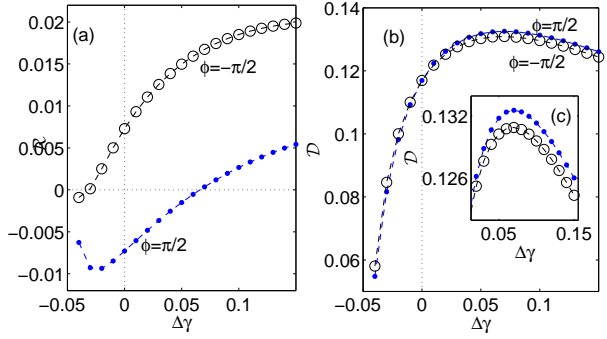}}} \caption{(a)-(b) Even
$\mathcal R$ and odd $\mathcal D$ conductance coefficients as a
function of the coupling asymmetry $\Delta\gamma=\gamma_R-\gamma_L$
with $\gamma_L=0.05$.
(c) Zoom over $\mathcal D$.
Other parameters are
$\epsilon_1=\epsilon_2=0.15$, $\gamma_P=0.1$, $\Delta\mu=0.4$ and
$\beta_a=50$.} \label{Fig5}
\end{figure}

\begin{figure}[t]
\hspace{1mm} {\hbox{\epsfxsize=80mm \epsffile{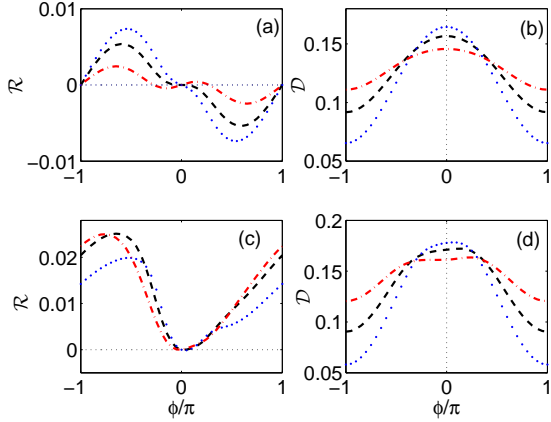}}}
\caption{Effect of the voltage probe coupling energy on even ($\mathcal R$)
and odd  ($\mathcal D$) conductance terms. (a)-(b) Spatially
symmetric system, $\gamma_L=\gamma_R=0.05$. (c)-(d) Spatially
asymmetric junction, $\gamma_L=0.05\neq\gamma_R=0.2$.
$\gamma_P=0.1$ (dot), $\gamma_P=0.2$ (dashed line) and
$\gamma_P=0.4$ (dashed-dotted). Light dotted lines represent
symmetry lines. Other parameters are $\Delta\mu=0.4$,
$\epsilon_1=\epsilon_2=0.15$, $\beta_a=50$.} \label{Fig6}
\end{figure}

\begin{figure}[htpb]
\hspace{1mm} 
{\hbox{\epsfxsize=80mm \epsffile{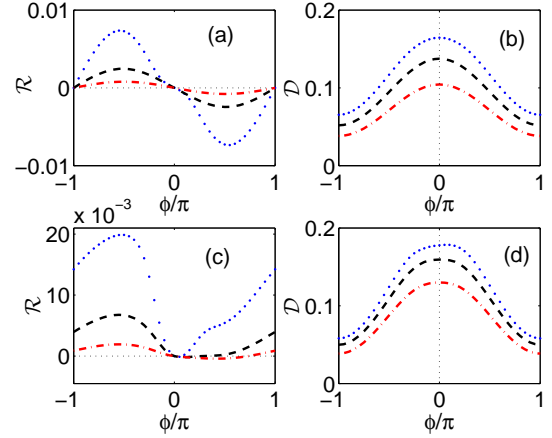}}} \caption{Temperature
dependence of even and odd conductance terms.
(a)-(b) Spatially symmetric system,
$\gamma_L=\gamma_R=0.05$. (c)-(d) Spatially asymmetric system,
$\gamma_L=0.05\neq \gamma_R=0.2$. In all panels $\beta_a=50$ (dots),
$\beta_a=10$ (dashed line) and $\beta_a=5$ (dashed-dotted line). The
light dotted lines mark symmetry lines. Other parameters are
$\Delta\mu=0.4$, $\gamma_P=0.1$ and  $\epsilon_1=\epsilon_2=0.15$.} \label{Fig3}
\end{figure}


\begin{figure}[ht]
\hspace{1mm} 
{\hbox{\epsfxsize=85mm \epsffile{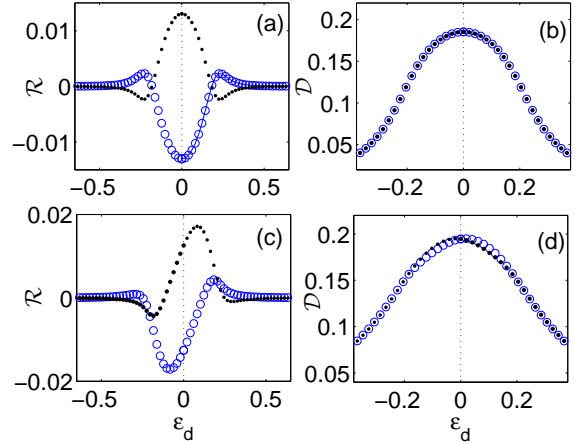}}} \caption{Magnetic field- gate voltage (MFGV)
symmetries.
(a)-(b) Even and odd conductance terms for a
spatially symmetric junction with $\gamma_L=\gamma_R=0.05$. (c)-(d)
Even and odd conductance terms for $\gamma_L=0.05$, $\gamma_R=0.2$,
demonstrating
 $\mathcal R(\epsilon_d,\phi)=-\mathcal R(-\epsilon_d,-\phi)$,
$\mathcal D(\epsilon_d,\phi)=\mathcal D(-\epsilon_d,-\phi)$.
In all cases $\phi=-\pi/4$ (small dots) and $\phi=\pi/4$ (empty
circle), $\Delta\mu=0.4$, $\gamma_{P}=0.1$ and $\beta_a=50$.}
\label{gate}
\end{figure} %


As the breakdown of the MF symmetries (\ref{eq:DI}) occurs under a
spatial asymmetry in the presence of inelastic effects, we study
next the role of $\gamma_P$, $\Delta \gamma\equiv \gamma_R-\gamma_L$
and the metals temperature $\beta_a^{-1}$ on these relations.

In Fig. \ref{Fig5} (a)-(b) we extract $\mathcal D$, and present it
along with $\mathcal R$ as a function of $\Delta \gamma$ We note
that the symmetry of $\mathcal R$ is feasibly broken with small
spatial asymmetry, while $\mathcal D$ is more robust.  The role of
the coupling strength $\gamma_P$ is considered in Fig. \ref{Fig6}.
First, in spatially symmetric systems we confirm again that the
MF relations (\ref{eq:DI}) are satisfied, and we note that as $\gamma_P$ increases, the
variation of $\mathcal D$ and $\mathcal R$ with phase is fading out.
Quite interestingly, the rectification contribution $\mathcal R$ may flip sign with $\gamma_P$,
for a range of phases.
(The sign of $\mathcal R$ reflects whether the total current has a larger
magnitude in the forward or backward bias polarity).
Second, when a spatial asymmetry is
introduced we note a strong breakdown of the MF phase symmetry
for $\mathcal R$, while the coefficient $\mathcal D$ still closely
follows the MF symmetry \cite{breakE2}.
Interestingly,  with increasing $\gamma_P$ the
variation of $\mathcal D$ with phase is washed out (panel d), but even conductance terms 
show a stronger alteration with $\phi$ (panel c). Thus, even and odd conductance terms respond distinctively to
decoherring and inelastic processes.

In Fig. \ref{Fig3} we consider the role of the reservoirs
temperatures on the conductance coefficients.
With increasing temperature a monotonic erosion of
the amplitude of all conductance terms with phase takes place. 
This should be contrasted to the non-monotonic
role of $\gamma_P$ on $\mathcal R$, as exposed in Fig. \ref{Fig5}.
%

Inspecting e.g., Fig. \ref{Fig3} we point out that
in our construction $\mathcal R(\phi=0)=0$, even
 in the presence of geometrical asymmetry, see discussion following Eq. (\ref{eq:RRR0}).
We recall that  many-body effects are presented here effectively, thus
this observation is not trivial given the common expectation that the combination
of many-body interactions and spatial asymmetry should bring in the current rectification effect
\cite{diodeH}. Indeed, 
extended models in which the system is
connected to the reservoirs indirectly, through ``linker" states,  present
rectification even at zero magnetic field, as long as both spatial
asymmetry and inelastic effects are introduced \cite{Malay}.

\begin{figure}[htbp]
\hspace{1mm} 
{\hbox{\epsfxsize=85mm \epsffile{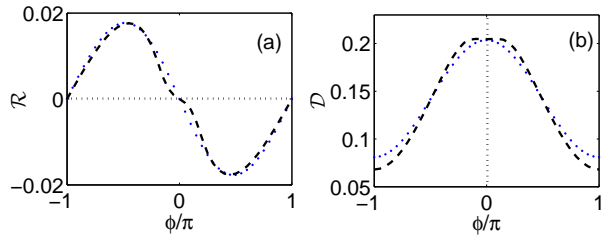}}} \caption{MF symmetries 
at the symmetric point $\epsilon_d=\epsilon_1=\epsilon_2=0$. (a) Even $\mathcal R$
(b) odd $\mathcal D$ conductance terms for spatially
symmetric $\gamma_L=\gamma_R=0.05$ (dots) and asymmetric situations
$\gamma_L=0.05$, $\gamma_R=0.2$ (dashed lines). Other parameters are
$\Delta\mu=0.4$, $\gamma_{P}=0.1$ and $\beta_a=50$.} \label{Fig7}
\end{figure}

\begin{figure}[thbp]
\hspace{-5mm}
{\hbox{\epsfxsize=85mm \epsffile{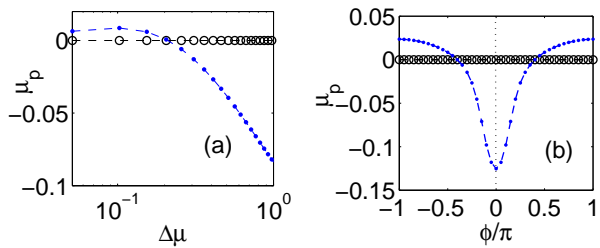}}}
 \caption{(a)-(b) Probe chemical potential at the symmetric point
for spatially symmetric $\gamma_L=\gamma_R=0.05$ (circles), and
asymmetric $\gamma_L=0.05\neq\gamma_R=0.2$ cases (dots). The lines
contain the overlapping $\phi=\pm \pi/4$ results. Other parameters
are $\Delta\mu=0.4$, $\beta_a=50$ and $\gamma_P=0.1$.} \label{Fig8}
\end{figure}

The scan of the current with $\epsilon_d=\epsilon_{1,2}$ is
presented in Fig. \ref{gate}.
When $\epsilon_d>\Delta \mu$, Onsager symmetry is practically respected
since the linear response limit is practiced, providing $\Delta
I\sim\mathcal R\sim0$. More significantly, this figure reveals the MFGV
symmetry (\ref{eq:CC}), valid irrespective of spatial asymmetries and many-body
(inelastic) effects.
This symmetry immediately implies that
at the so-called ``symmetric point", when
$\epsilon_d=0$ (set at the Fermi energy),  $\mathcal
R(\phi)$ is an odd function of the magnetic flux irrespective of spatial asymmetries.
This behavior is
displayed in Fig. \ref{Fig7}.
Also note that at the symmetric point the probe chemical potential does not manifest a linear response limit:
It is identically zero for symmetric setups, and it  satisfies
$\mu_P(\phi)=\mu_P(-\phi)$ far from equilibrium for setups
with a broken inversion symmetry, see Fig. \ref{Fig8}.
%

\subsection{Nonlinear transport  with non-dissipative inelastic effects}

\begin{figure}[t]
\hspace{1mm} 
{\hbox{\epsfxsize=70mm \epsffile{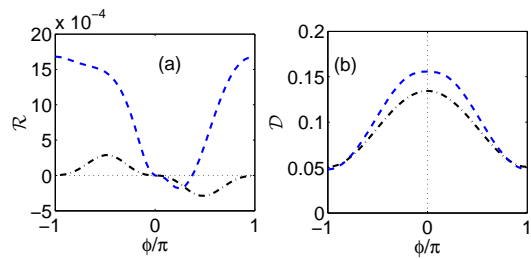}}}
 \caption{Voltage-temperature probe.
(a) $\mathcal R$ and (b) $\mathcal D$ in spatially symmetric case (dashed-dotted lines) $\gamma_{L,R}=0.05$
 and asymmetric setups (dashed lines) $\gamma_L=0.05\neq \gamma_R=0.2$.
$\beta_a=10$, $\gamma_P=0.1$, $\epsilon_{1,2}=0.15$. 
}\label{probe3}
\end{figure}

The simulations presented throughout Figs. \ref{Fig1}-\ref{Fig8} were obtained under
the voltage probe condition, thus heat
dissipation takes place at the probe. In Fig. \ref{probe3} we show that the breakup of the MF
symmetries occurs in spatially asymmetric setups under the more restrictive voltage-temperature probe, when only
{\it non-dissipative} inelastic effects are allowed.
We again note that the breakdown of the phase symmetry of $\mathcal D$ is 
small, one order of magnitude below the variation in $\mathcal R$.




\section{Summary}
\label{summary}

We have studied the role of elastic and inelastic scattering effects
on magnetic field symmetries of nonlinear conductance terms
using B\"uttiker's probe technique. 
For spatially symmetric junctions we proved the validity of the MF symmetries
$\mathcal D(\phi)=\mathcal D(-\phi)$ and $\mathcal
R(\phi)=-\mathcal R(-\phi)$,
though many-body inelastic effects, introduced via the probe, are
asymmetric in magnetic flux. 
We demonstrated the 
breakdown of these MF symmetries when the junction has a
left-right asymmetry, in the presence of inelastic effects.
Using a double-dot AB interferometer model we showed that 
it respects more general MFGV symmetry relations,
$\mathcal R(\epsilon_d,\phi)=-\mathcal R(-\epsilon_d,-\phi)$ and
$\mathcal D(\epsilon_d,\phi)=\mathcal D(-\epsilon_d,-\phi)$.

The rectification effect, of fundamental and practical interest, is
realized by combining many-body interactions with a broken
symmetry: broken spatial inversion symmetry or a broken time
reversal symmetry. Rectifiers of the first type have been
extensively investigated theoretically and experimentally, including
electronic rectifiers, thermal rectifiers \cite{diodeH}
and acoustic rectifiers \cite{diodeA}.
In parallel, optical and spin rectifiers were designed
based on a broken time reversal symmetry, recently realized e.g. 
by engineering parity-time meta-materials \cite{diodeO}.

The model system investigated in Secs.\ref{violation}-\ref{simulation}, the double-dot AB
junction, offers a feasible setup for devising broken-time
reversal rectifiers:
We found that $\mathcal R\neq 0$ when two
conditions are simultaneously met: the magnetic flux obeys
$\phi\neq 2\pi n$, $n$ is an integer, and the probe introduces inelastic effects.
However, when time-reversal symmetry is maintained
(when the flux obeys $\phi=2n\pi$) our model does
not provide the rectification effect even if the spatial mirror symmetry is
broken. The technical reason is that in our minimal construction
both dots are coupled to the $L$ and $R$ metals directly, with an energy independent hybridization
constant. In extended models when the ring is coupled indirectly to
the $L$ and $R$ metals, through a spacer state, geometrical
rectification may develop even in the absence of a magnetic flux.

It is of interest to verify the results of this work by adopting a
microscopic model with genuine many-body interactions \cite{mic-ee,UedaD,KuboD}, by modeling a quantum
point contact \cite{QPC1,QPC2} or an equilibrated phonon bath,
exchanging energy with the junction's electronic degrees of freedom.
This could be done by extending numerical and analytic
studies, e.g., Refs. \cite{Hod, Hod2,Ora}, to the nonlinear regime.
It is also of interest to obtain the MF and the MFGV symmetries 
 using a full-counting statistics approach \cite{Saito,ButtikerFCS}.

Future studies will be devoted to the analysis 
of the thermoelectric effect under broken time-reversal symmetry 
\cite{Casati1,Casati2, Seifert,Ora,Lopez,Serra,Whitney} in the far-from-equilibrium regime,
and to the study of quantum transport, far-from-equilibrium, in networks with 
broken time-reversal symmetry \cite{TRS}.

\acknowledgments DS acknowledges support from the Natural Science and Engineering Research 
Council of Canada.
The work of SB has been supported by the Early Research Award of DS.
The authors acknowledge R. H\"artle for useful comments.


\renewcommand{\theequation}{A\arabic{equation}}
\setcounter{equation}{0}  
\section*{Appendix A: Charge conjugation symmetries for a double-dot AB interferometer}

We derive the MFGV symmetry relations by considering
a double-dot interferometer model which does not necessarily acquire a spatial symmetry.
Given the Hamiltonian (\ref{eq:H})-(\ref{eq:HSP}), we introduce a charge
conjugation operator $\mathcal C$ that acts to replace
an electron by a hole \cite{OraC},
\bea
\mathcal C (\epsilon_d)&=&-\epsilon_d, \,\,\,\  \mathcal C (\epsilon_k)=-\epsilon_k, 
\nonumber\\
\mathcal C (v_{n,j}e^{i\phi_n})&=& -v^*_{n,j}e^{-i\phi_n}, \,\,\,\
\mathcal C (\langle a^\dagger a\rangle) = 1-  \langle a^\dagger a\rangle.
\nonumber\\
\label{eq:AppTr}
\eea
Here $a^{\dagger}$ and $a$ are fermionic creation and annihilation operators, respectively.
First, we need to find  what symmetries does the 
probe chemical potential obey.
We achieve this by studying the probe condition (\ref{eq:fpp}) as is, under reversed bias voltage, 
and under charge conjugation. Note that in our model the transmission function satisfies 
$\mathcal T(\epsilon,\epsilon_d,\phi)=\mathcal T(-\epsilon,-\epsilon_d,-\phi)$,
i.e. it is invariant under charge conjugation. The probe condition fulfills
\begin{widetext}
\bea
&&
\int d\epsilon (\mathcal T_{P,L}+\mathcal T_{P,R})f_ P(\mu_P(\epsilon_d,\phi,\mu_L,\mu_R))= \int d\epsilon (\mathcal T_{L,P}f_L +\mathcal
T_{R,P}f_R)
\label{eq:fppGa}
\\
&&\int d\epsilon(\mathcal T_{P,L}+\mathcal T_{P,R}) f_P(\mu_P(\epsilon_d,\phi,\mu_R,\mu_L))=\int d\epsilon\left(\mathcal T_{L,P}f_R+\mathcal
T_{R,P}f_L\right).
\label{eq:fppGb}
\\
&&
\int d\epsilon(\mathcal{T}_{P,L}+\mathcal{T}_{P,R}) [1-f_P( -\mu_P(-\epsilon_d,-\phi,\mu_L,\mu_R))]
= \int d\epsilon (   \mathcal{T}_{L,P}[1-f_R] + \mathcal
    {T}_{R,P}[1-f_L])
\label{eq:fppG}
\eea
\end{widetext}
For clarity, we explicitly noted
the dependence of $f_P$ on the chemical potential $\mu_P$, itself obtained
given the set of parameters $\epsilon_d$, $\phi$, and $\mu_{L,R}$.
The last relation [Eq. (\ref{eq:fppG})]  has been derived by noting that
\bea
f_{\nu}(-\epsilon,\mu_{\nu})&=&[e^{\beta(-\epsilon-\mu_{\nu})}+1]^{-1}
\nonumber\\
&=&1-f_{\nu}(\epsilon,-\mu_{\nu}). 
\eea
Since we use the convention $\mu_L=-\mu_R$, we conclude that 
$f_{L}(-\epsilon,\mu_{L})=1-f_R(\epsilon,\mu_R)$.
Next we omit the direct reference to the energy $\epsilon$ within $\mathcal T$ and 
the Fermi functions. 
Also, we do not write the phase $\phi$ in the transmission function, 
unless necessary to eliminate confusion.
Equation (\ref{eq:fppG}) now reduces to
\bea
&& \int d\epsilon(\mathcal{T}_{P,L}+\mathcal{T}_{P,R}) f_P( -\mu_P(-\epsilon_d,-\phi,\mu_L,\mu_R))
\nonumber\\
&&= \int d\epsilon (   \mathcal{T}_{L,P}f_R + \mathcal{T}_{R,P}f_L).
\eea
Comparing this to  Eq. (\ref{eq:fppGb}), we immediately note that
\bea
f_P( -\mu_P(-\epsilon_d,-\phi,\mu_L,\mu_R))=f_P(  \mu_P(\epsilon_d,\phi,\mu_R,\mu_L).
\nonumber\\
\label{eq:fppss}
\eea
Since all reservoirs are maintained at $T_a$ this implies that
\bea
\mu_P(\epsilon_d,\phi,\mu_L,\mu_R)=- \mu_P(-\epsilon_d,-\phi,\mu_R,\mu_L).
\eea
%
This relation should be compared to Eq. (\ref{eq:fpS}),
valid for spatially symmetric systems. 
We now utilize Eq. (\ref{eq:fppss}) and derive the symmetry relations for
$\mathcal R$ and $\mathcal D$. 
We recall the explicit expressions for these measures, see definitions in Sec. \ref{measure},
\begin{widetext}
\bea
\mathcal R&=&\frac{1}{2}\int d\epsilon
\mathcal T_{P,L}\left[ f_L+f_R-f_P(\mu_P(\epsilon_d,\phi,\mu_L,\mu_R))-
 f_P(\mu_P(\epsilon_d,\phi,\mu_R,\mu_L)) \right]
\nonumber\\
\mathcal D&=&\frac{1}{2}\int d\epsilon
\Big\{(\mathcal T_{L,R}+\mathcal T_{L,P}+\mathcal T_{R,L})( f_L-f_R)
-\mathcal T_{P,L}
\left[ f_P(\mu_P(\epsilon_d,\phi,\mu_L,\mu_R))-f_P(\mu_P(\epsilon_d,\phi,\mu_R,\mu_L)\right] \Big\}
\eea
\end{widetext}
The charge-conjugated expressions satisfy
\begin{widetext}
\bea
\mathcal C(\mathcal {R})&=&\frac{1}{2}\int d\epsilon
\mathcal {T}_{P,L}\left\{ [1-f_R]+ [1-f_L]-[1-f_P(-\mu_P(-\epsilon_d,-\phi,\mu_L,\mu_R))] -[1- f_P(-\mu_P(-\epsilon_d,-\phi,\mu_R,\mu_L)] \right\}
\nonumber\\
&=&-\frac{1}{2}\int d\epsilon
\mathcal {T}_{P,L}\left\{ f_R+f_L- f_P(-\mu_P(-\epsilon_d,-\phi,\mu_L,\mu_R))-f_P( -\mu_P(-\epsilon_d,-\phi,\mu_R,\mu_L)) \right\}
\nonumber\\
&=& -\mathcal R
\eea
The last equality is reached by using Eq. (\ref{eq:fppss}). Similarly, we obtain 
the symmetry of odd conductance terms as
\bea
\mathcal C( \mathcal{D}) &=&\frac{1}{2}\int d\epsilon
\Big\{(\mathcal {T}_{L,R}+\mathcal { T}_{L,P}+\mathcal {T}_{R,L})
( 1-f_R- 1+f_L)
\nonumber\\
&-&\mathcal { T}_{P,L}
\left[ 1-f_P(- \mu_P(-\epsilon_d,-\phi,\mu_L,\mu_R))
-1+f_P(-\mu_P(-\epsilon_d,-\phi,\mu_R,\mu_L)) \right] \Big\}
\nonumber\\
&=&
\frac{1}{2}\int d\epsilon
\Big\{(\mathcal { T}_{L,R}+\mathcal { T}_{L,P}+\mathcal { T}_{R,L})
(f_L-f_R)
+\mathcal { T}_{P,L}
\left[ f_P(- \mu_P(-\epsilon_d,-\phi,\mu_L,\mu_R))
-f_P(-\mu_P(-\epsilon_d,-\phi,\mu_R,\mu_L)) \right] \Big\}
\nonumber\\
&=& \mathcal D
\label{eq:ACCC}
\eea
\end{widetext}
In the double-dot interferometer model these relations translate to the MFGV
symmetries,
\bea
\mathcal R(\epsilon_d,\phi)&=&-\mathcal R(-\epsilon_d,-\phi), 
\nonumber\\
\mathcal D(\epsilon_d,\phi)&=&\mathcal D(-\epsilon_d,-\phi).
\eea
For compactness, the derivation above has been performed assuming $\epsilon_d=\epsilon_1=\epsilon_2$.
It is trivial to note that our results are valid beyond degeneracy.


\renewcommand{\theequation}{B\arabic{equation}}
\setcounter{equation}{0}  
\section*{Appendix B: MF symmetries of nonlinear heat transport}

In this Appendix we study symmetry relations of the electronic heat
current under nonzero magnetic flux and a temperature bias $T_L\neq
T_R$, in the absence of a potential bias, $\mu_a=\mu_L=\mu_R=\mu_P$.
Using the temperature probe (\ref{eq:Tprobe}), we demand that heat dissipation at the probe diminishes, 
$Q_P=0$, but allow for charge dissipation. 
We now express the $L$ to $R$ heat current $Q_{\Delta T}\equiv
Q_L=-Q_R$ in powers of the temperature bias as
\bea Q_{\Delta T}(\phi)=K_1(\phi) \Delta T + K_2(\phi) (\Delta T)^2
+ K_3(\phi)(\Delta T)^3 +... \label{eq:QIIT},
\nonumber\\
\eea
where $K_{n>1}$ are the nonlinear conductance coefficients. These
coefficients depend on the junction parameters: energy,
hybridization and possibly the temperature $T_a=(T_L+T_R)/2$. We
define next symmetry measures for the heat current $Q_{\Delta T}$,
parallel to Eqs. (\ref{eq:DeltaIphi})-(\ref{eq:R}). First, we collect 
even conductance terms into $\mathcal R_{\Delta T}$,
\bea &&\mathcal R_{\Delta T}(\phi)\equiv \frac{1}{2}[Q(\phi)+\bar
Q(\phi)]
\nonumber\\
&& = K_2(\phi) (\Delta T)^2  + K_4(\phi) (\Delta T)^4+ ...
\nonumber\\
&&= \mathcal  \int \frac{\mathcal T_{P,L}-\mathcal
T_{P,R}}{4}(f_L+f_R-f_P(\phi)-\bar f_P(\phi))
(\epsilon-\mu_a)d\epsilon
\nonumber\\
\label{eq:QR} \eea
Here $\bar Q$ is defined as the heat current obtained upon
interchanging the temperatures of the $L$ and $R$ terminals. We also study the behavior of
odd conductance terms,
\bea \mathcal D_{\Delta T}(\phi)\equiv  K_1(\phi) \Delta T  +
K_3(\phi) (\Delta T)^3+ ... \label{eq:QD} \eea
In the absence of the probe and in the linear response limit the heat
current satisfies an even phase symmetry, $Q_{\Delta
T}(\phi)=Q_{\Delta T}(-\phi)$. Deviations from this symmetry are
collected into the  measure
 \bea \Delta Q_{\Delta T}&=& \frac{1}{2}\left[Q_{\Delta
T}(\phi)-Q_{\Delta T}(-\phi)\right] \nonumber\\ &=& \frac{1}{2} \int
\left[ \mathcal T_{L,R} - \mathcal T_{R,L}\right] (\epsilon-\mu_a)
f_R d\epsilon
\nonumber\\
&+& \frac{1}{2} \int \left[ \mathcal T_{L,P}f_P(-\phi) -
\mathcal T_{P,L} f_P(\phi) \right](\epsilon-\mu_a)d\epsilon.
\nonumber\\ \label{eq:QDeltaI} \eea
%


{\it Linear response regime.}
We repeat the derivation of Sec. \ref{dephasing}, 
and find that in the linear response regime, $T_{L}=T_a+\delta
T_{L}$, $T_{R}=T_a+\delta T_{R}$, the probe temperature
 $T_P=T_a+\delta T_P$ obeys
\bea &&\delta T_P(\phi) = \frac{\int d
\epsilon\left(\frac{-\partial f_a}{\partial
 \epsilon}\right)\frac{(\epsilon-\mu_a)^2}{T_a} \left( \delta
T_L\mathcal T_{L,P}+\delta T_R\mathcal T_{R,P}\right)} {\int d
\epsilon\left(\frac{-\partial f_a}{\partial
\epsilon}\right)\frac{(\epsilon-\mu_a)^2}{T_a} \left( \mathcal
T_{P,L}+\mathcal T_{P,R}\right)} \nonumber\\\eea
Using this relation, one can readily repeat the steps in Sec.
\ref{dephasing} and prove that $\mathcal R_{\Delta T}=0$, thus
$Q_{\Delta T}(\phi)=\mathcal D_{\Delta T}(\phi)=Q_{\Delta
T}(-\phi)$.

{\it Symmetry relations far-from-equilibrium.} We discuss here
symmetry relations for spatially symmetric junctions. We adapt the
temperature probe condition (\ref{eq:Tprobe}) to three situations.
First, the standard expression is given by
\bea&&\int d\epsilon (\mathcal T_{P,L}+\mathcal T_{P,R})f_
P(\phi) (\epsilon-\mu_a) \nonumber\\ &&= \int d\epsilon
(\mathcal T_{L,P}f_L +\mathcal T_{R,P}f_R)(\epsilon-\mu_a). \eea
We reverse the magnetic phase and get
\bea &&\int d\epsilon(\mathcal T_{L,P}+\mathcal
T_{R,P})f_P(-\phi)(\epsilon-\mu_a) 
\nonumber\\ && = \int
d\epsilon( \mathcal T_{P,L}f_L +\mathcal
T_{P,R}f_R)(\epsilon-\mu_a).
\label{eq:Qfpp2}
\eea
Similarly, when interchanging the temperatures $T_L$ and $T_R$ we
look for the probe distribution  $\bar f_P$ which satisfies
 \bea &&\int d\epsilon(\mathcal T_{P,L}+\mathcal T_{P,R})\bar
f_P(\phi)(\epsilon-\mu_a) \nonumber\\ &&=\int
d\epsilon\left(\mathcal T_{L,P}f_R+\mathcal T_{R,P}f_L\right)
(\epsilon- \mu_a). \label{eq:Qfpp3} \eea
Note that  $f_P(\phi)$, $f_P(-\phi)$ and $\bar f_P(\phi)$ are required to 
follow a Fermi-Dirac form. The temperature $\beta_{P}$ should be obtained
so as to satisfy the probe condition.
%
If the junction is left-right symmetric, the mirror symmetry
 $\mathcal T_{P,L}(\phi)=\mathcal T_{R,P}(\phi)$ applies. We
use this relation in Eqs. (\ref{eq:Qfpp2}) and (\ref{eq:Qfpp3}) and conclude that
the probe distribution obeys
\bea 
\bar 
f_P(\phi)=f_P(-\phi). 
\label{eq:QfpS} \eea
This directly implies that ($\mu_a=\mu_L=\mu_R=\mu_P$)
\bea 
\bar \beta_P(\phi)=\beta_P(-\phi). 
\label{eq:bfpS} 
\eea %
Note that  $\beta_P(\phi)$ does not
need to obey any particular magnetic phase symmetry. The deviation from
phase rigidity, Eq. (\ref{eq:QDeltaI}), can be expressed using
the heat current flowing into the $R$ terminal,
\bea \Delta Q_{\Delta T}&=& \frac{1}{2}\int(\epsilon-\mu_a)
[(\mathcal T_{L,R}-\mathcal T_{R,L}) f_L
\nonumber\\
&-&\mathcal T_{R,P}f_P(-\phi) +\mathcal T_{P,R}f_P(\phi)]d\epsilon.
\label{eq:QDeltaI2} \eea
We define $\Delta Q_{\Delta T}$ as the average of Eqs.
(\ref{eq:QDeltaI}) and (\ref{eq:QDeltaI2}), 
\bea \Delta Q_{\Delta T}&=&\frac{1}{4}\int
d\epsilon(\epsilon-\mu_a)\Big[(\mathcal T_{L,R}-\mathcal
T_{R,L})(f_L+f_R)
\nonumber\\
&+&(\mathcal T_{L,P}-\mathcal T_{R,P})f_P(-\phi) +(\mathcal
T_{P,R}-\mathcal T_{P,L})f_P(\phi)\Big]. \nonumber \eea
Using the identities $\mathcal T_{L,R}-\mathcal T_{R,L} =
\mathcal T_{P,L}-\mathcal T_{L,P}$ and  $\mathcal T_{P,L}=\mathcal T_{R,P}$,
the latter is valid in geometrically symmetric junctions, we get
\bea && \Delta Q_{\Delta T}(\phi) \nonumber\\= && \frac{1}{4}\int
(\mathcal T_{P,L}-\mathcal T_{P,R})(f_L+f_R-f_P(\phi)-\bar
f_P(\phi))(\epsilon-\mu_a) d\epsilon
\nonumber\\
&&= \mathcal R_{\Delta T}(\phi)=- \mathcal R_{\Delta T}(-\phi).
\eea
%
This concludes our derivation that under a temperature bias even (odd) heat
conductance coefficients satisfy an odd (even) magnetic field symmetry,
\bea \mathcal R_{\Delta T}(\phi)&=&-\mathcal R_{\Delta
T}(-\phi)=\Delta Q_{\Delta T}(\phi),\,\,\,\,\, \nonumber\\
\mathcal D_{\Delta T}(\phi)&=&\mathcal D_{\Delta T}(-\phi),
\label{eq:ADI} \eea
as long as the junction acquires a spatial mirror symmetry.

\begin{figure}[htbp]
\hspace{1mm} {\hbox{\epsfxsize=80mm \epsffile{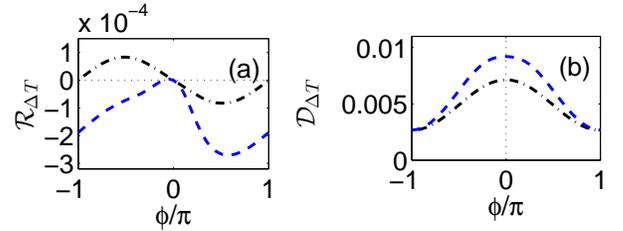}}}
\caption{Magnetic field symmetries of (a) even and (b) odd
electronic heat conductance terms.
Spatially symmetric system (dashed dotted),
$\gamma_L=\gamma_R=0.05$.
Spatially asymmetric junction (dashed),
$\gamma_L=0.05\neq\gamma_R=0.2$. Light dotted lines represent the
symmetry lines. Other parameters are  $T_L=0.15$,  $T_R=0.05$,
$\epsilon_1=\epsilon_2=0.15$, $\gamma_P=0.1$, $\mu_a=\mu_L=\mu_R=\mu_P=0$.
} \label{FigA}
\end{figure}


We adopt the double-dot model (\ref{eq:H}) presented in Sec. 
\ref{violation}
and study its heat current behavior. In the absence of the
probe, assuming for simplicity degeneracy and spatial symmetry,
$\gamma/2=\gamma_{L,R}$, we obtain
\begin{widetext}
\bea Q_L(\phi)=
\int
d\epsilon (\epsilon-\mu_a) \frac{\gamma^2(\epsilon-\epsilon_d)^2\cos^2\frac{\phi}{2}}
{\left[(\epsilon-\epsilon_d)^2-\frac{\gamma^2}{4}\sin^2\frac{\phi}{2}\right]^2+\gamma^2(\epsilon-\epsilon_d)^2}
[f_L(\epsilon)-f_R(\epsilon)], \nonumber \eea
\end{widetext}
satisfying the Onsager symmetry. Fig. \ref{FigA} displays the
MF symmetries, and their violation, 
Deviations from phase
symmetry for $\mathcal D_{\Delta T}$ are small, of the order of
$10^{-5}$.




\begin{thebibliography}{32}
\bibitem{Persistent}
Persistent currents exist in at thermal equilibrium due
to the absence of time-reversal symmetry.
In our work here we consider only net charge transport. However,
since persistent currents are antisymmetric in the magnetic field,
we could add them to the defintion of $\mathcal R(\phi)$, and our
observations would be still valid.
\end{thebibliography}
\end{document}